\documentclass[]{article}

\usepackage{amsmath, amsthm, amssymb, amsfonts}
\usepackage{mathtools}
\usepackage{hyperref}
\usepackage{graphicx}
\usepackage{dsfont}
\usepackage{fullpage}
\usepackage{color}
\usepackage{soul}
\usepackage[title]{appendix}
\usepackage{tikz}
\usetikzlibrary{patterns}
\usetikzlibrary{shapes.multipart}
\usetikzlibrary{arrows}

\theoremstyle{definition}

\definecolor{labelkey}{cmyk}{.4,.2,0,0}

\newcommand{\be}{\begin{equation}}
\newcommand{\ee}{\end{equation}}
\newcommand{\bea}{\begin{eqnarray}}
\newcommand{\eea}{\end{eqnarray}}

\usepackage{titlesec}
\titleformat{\section}{\large\bf}{\thesection}{1em}{}
\titleformat{\subsection}[runin]{\bf}{\thesubsection}{1em}{}[.]

\usepackage{authblk}
\author[1]{Yan Fyodorov}
\author[2]{Pierre Le Doussal}
\author[1]{Alexander Ossipov}
\affil[1]{\normalsize  Department of Mathematics, King's College London, Strand, London, WC2R 2LS, UK}
\affil[2]{\normalsize Laboratoire de Physique de l'\'Ecole Normale Sup\'erieure, CNRS, ENS and 
Universit\'e PSL, Sorbonne Universit\'e, Universit\'e Paris Cit\'e, 75005 Paris, France}

\title{\bf \large Large deviations of spectral determinants of matrix-valued random Schr\"odinger operators
and Dyson Brownian motion in cubic potentials}
\date{}

\begin{document}

\maketitle

\begin{abstract}
We study the moments of
$\overline{|\det(H-E)|^q}$ and the associated large deviations of $\log |\det(H-E)|$
where $H$ are random matrix operators involving Laplace operators and random potentials.
This includes as a special case Hessians of random elastic manifolds at a generic energy configuration.
In one dimension $d=1$ these are $N \times N$ matrix valued random Schr\"odinger operators
and $\log | \det(H-E) |$ is the sum of the $N$ associated Lyapunov exponents.
Using a mapping to a stochastic matrix Ricatti equation we make a connection
between the spectral properties of these operators and the total $N$ particle current of a Dyson Brownian
motion (DBM) in a cubic potential. The latter model was studied by
Allez and Dumaz \cite{AD15} who showed that for $N=+\infty$ it exhibits a sharp transition between
a phase with non-zero current and a confined (zero current) phase.
We compute the barrier-crossing probability of the DBM at large but finite $N$,
which gives an estimate of the exponential tail of the average density of states
 of a matrix Schrodinger operator below the edge of its spectrum. The barrier behaves as $\sim N (-E)^{3/2}$ at large negative energy and vanishes
as $\sim N(E^*-E)^{5/4}$ near the edge.
For $q=1$ the present work provides an independent derivation of the total complexity of stationary points
for an elastic string embedded
in $N$ dimension in presence of disorder.
\end{abstract}


\section{Introduction}

Matrix-valued Schr\"odinger
operators involving Laplace operators (or their discrete lattice analogues) and random potentials appear in many contexts.
Originally they have been used as a model for studying Anderson localization in quasi-one dimensional multichannel systems (wires or strips) in
classical works by Dorokhov \cite{Dor88} and Mello-Pereyra-Kumar \cite{MPK88}. Those works introduced the powerful DMPK approach to
disordered conductors \cite{B97}, with mathematical aspects of the approach and its relation to underlying operators remaining an active subject, see \cite{BdR10,VV14} and refs therein.
More recently the same operators appeared naturally as Hessians of elastic disordered systems \cite{UsHessianManifold,UsManifComplexity,McKenna1,McKenna2}.
Those studies, extending earlier results available for scalar (strictly one-dimensional) Schr\"odinger operators \cite{FLRT},  attracted attention to the statistics of spectral determinants of their matrix-valued analogues,  which turned out to play central role in counting abundant mechanical equilibria.
The latter are known to be responsible for glassy properties of disordered elastic systems, see e.g. \cite{TGPLDBragg2}.

In this paper we will be interested in two versions of this class of random operators:

\begin{itemize}

\item

{\it Discrete operators} defined by the square matrix of size $N M$
\be \label{discretemodel}
{\cal H}_{ix,jy} = \delta_{ij} (\mu \mathbf{1} - t \Delta)_{xy} + W_{ij}(x) \delta_{xy}
\ee
where $x,y$ are integer labels in $1,\dots,M$ and the matrices $W(x)$ are $N \times N$ and real symmetric. For applications below,
$\Delta_{xy}$ is the discrete Laplacian in dimension $d$ with $M=L^d$ where $L$ is the linear size of the system. ${\cal H}$ can thus be represented a block matrix (banded in $d=1$) with blocks of size $N \times N$. The matrices $W(x)$ are random and represent the disorder.
They are centered Gaussian distributed,
with correlations given by
\bea \label{correlatorW}
&& \overline{W_{ij}(x) W_{kl}(x')} = \frac{J^2}{N} (\delta_{ik} \delta_{jl} + \delta_{il} \delta_{jk} + c \delta_{ij} \delta_{kl} )
\delta_{x,x'}
\eea
where the parameter $J$ controls the strength of the disorder and $c\ge 0$.
Equivalently they can be written as
\be
W(x) = J \, ( H(x) + \xi(x) I )
\ee
where the $H(x)$ form a set of $M$ GOE($N$) matrices, independent for different $x$ distributed with $P(H) \sim e^{- \frac{N}{4} ({\rm Tr} H)^2}$ (a normalisation such that at large $N$ the spectrum of $H$ is a semi-circle supported
in $[-2,2]$). The variables $\xi(x)$ are i.i.d. Gaussian with PDF
\be \label{Pxi}
P(\xi) \sim e^{- N \sum_x \frac{\xi(x)^2}{2 c}} \, ,
\ee
and are independent of the $H(x)$.

It is convenient to write the matrix ${\cal H}$ in \eqref{discretemodel} as a sum
\be \label{Hsum}
{\cal H} = K + X + \mu I \quad , \quad K_{ix,jy}=  - t \Delta_{xy}  \delta_{ij} + J  H(x)_{ij} \delta_{xy}
\quad , \quad X_{ix,jy}=  J \xi(x) \delta_{ij} \delta_{xy}
\ee
For example for $d=1$, $L=M=2$
\be
X =  \begin{pmatrix} J \xi(1) I_N & 0  \\
0 & J \xi(2) I_N
\end{pmatrix}
~ , ~ K = \begin{pmatrix}
 J H(1) + 2  t I_N & -  2 t I_N  \\
- 2 t  I_N &  J H(2) + 2  t I_N
\end{pmatrix}
\ee
For larger sizes, see the Figure on page 180 in \cite{UsHessianManifold}.

\item

{\it Continuous matrix valued Schr\"odinger operators} in $d=1$ of the type
\be \label{contmodel}
{\cal H}_{ij} = (-  \tilde t \frac{d^2}{d \tau^2} + \tilde \mu ) \delta_{ij} + \tilde W_{ij}(\tau)
\ee
which act on vector functions $\psi(\tau)=(\psi_1(\tau), \dots \psi_N(\tau))$
of the continuous variable $\tau \in  [0,L]$,
where $L$ is the sample size. Here $\tilde W(\tau)$ is a $N \times N$ Gaussian white noise random matrix process (understood below in Ito convention) with correlations
\bea
&& \overline{W_{ij}(\tau) W_{kl}(\tau')} = \frac{\tilde J^2}{N} (\delta_{ik} \delta_{jl} + \delta_{il} \delta_{jk} + c \delta_{ij} \delta_{kl} )
\delta(\tau-\tau')
\eea
Again this is equivalent to writing $\tilde W(\tau) = \tilde J ( \tilde H(\tau) + \tilde \xi(\tau) )$
where $\tilde \xi(\tau)$ is $\sqrt{c/N}$ times a standard white noise process, and $\tilde H(\tau)_{ij} = \frac{\eta_{ij}(\tau) + \eta_{ji}(\tau)}{\sqrt{2 N}}$ where all $N^2$, $\eta(\tau)_{ij}$ are independent standard white noises. Note that
the continuum model can be obtained from the discrete one in $d=1$, see details below.
Note that determinants of such operators have been studied recently
\cite{Gruzberg1}
in
the context of instanton calculations for quasi-1D conductors.\\

\end{itemize}

Both operators arise as the Hessian matrix associated to the energy functional of an
elastic manifold of internal dimension $d$ in a random potential,
embedded in dimension $d+N$, and taken at a generic manifold configuration. The first operator is associated to
a discrete model of a disordered elastic manifold of arbitrary $d$ \cite{UsHessianManifold,UsManifComplexity},
a generalization of the similar  $d=0$ problem for a single particle in a disordered potential \cite{Fyo04,FyoNad12,UsHessToy}.
The second operator arises for the continuum limit of the same problem in $d=1$, and is associated to
a directed elastic line embedded in dimension $1+N$, as studied in \cite{FLRT} for $N=1$.
In the discrete model the manifold is parameterized by a $N$-component field ${\bf u}(x) \in \mathbb{R}^N$,
where $x$ spans an internal space $x \in \Omega$ on a discrete lattice $\Omega \subset \mathbb{Z}^d$.
The associated energy functional  \cite{MezPar91,MP2} 
\begin{equation}\label{landscape}
{\sf H}[{\bf u}]= \frac{1}{2} \sum_{x,y} {\bf u}(x) \cdot (\mu \mathbf{1} - t \Delta)_{xy} \cdot {\bf u}(y)
+ \sum_{x} V(\mathbf{u}(x),x)
\end{equation}
is the sum of an elastic energy described by the Laplacian matrix
$- t \Delta_{xy}$, $t>0$, a quadratic confining
energy controlled by the curvature $\mu>0$ and
a centered Gaussian random potential $V({\bf u},x)$
with covariance
\begin{equation}\label{cov}
\overline{ V(\mathbf{u}_1,x_1) V(\mathbf{u}_2,x_2) } = N\: B\left(\frac{(\mathbf{u}_1-\mathbf{u}_2)^2}{N}\right) \delta_{x_1,x_2}
\end{equation}
parametrized by a function $B(z)$. The model has long history of research in physical literature, with emphasis on its glassy behaviour, see e.g. \cite{LeDWie04}
and references therein, and in $N\to \infty$ limit and low temperatures is controlled by the effects of replica symmetry breaking, see e.g. \cite{MezPar91,MP2,LeDWie03,LeDWie08} and more recently in \cite{UsHessianManifold,BAK24a,BAK24b}.
In such a model the Hessian matrix around a generic configuration ${\bf u}(x)$  reads
\bea \label{Hessian}
{\cal K}_{ix,jy}[{\bf u}] &=& \frac{\partial ^2}{\partial u_{i}(x) \partial u_j(y)}{\sf H}[{\bf u}] = \delta_{ij} (\mu \mathbf{1} - t \Delta)_{xy} + \delta_{xy} \frac{\partial ^2}{\partial u_{i}\partial u_j}V({\bf u}(x),x)\,.
\nonumber
\eea
Because of the form chosen for correlations in \eqref{cov} the
distribution of the Hessian matrix is independent of the choice of the configuration
${\bf u}(x)$, hence we can choose ${\bf u}(x)=0$.
As shown in \cite{UsHessianManifold} the matrix ${\cal K}[0]$ has the same probability distribution as the matrix
${\cal H}$ in \eqref{discretemodel} with the parameter $J^2 = 4 B''(0)$ and $c=1$.
\\

The aim of the present paper is to study the moments of the modulus of the spectral determinant
\be  \label{Yq}
Y_q = \overline{ |\det ({\cal H}- E)|^q },
\ee
where $q$ is real and $q > -1$ \footnote{Note that for $q \leq -1$ one needs an additional regularization for the moments to be finite \cite{FyoKeating03}, hence we will not consider this case here.}.
One finds that at large $L$ they grow as
\be  \label{defSigmaq}
\tilde Y_q = \frac{Y_q}{Y_0^q} \sim e^{N L^d  \Sigma_q},
\ee
where it is convenient
to define the reduced moments $\tilde Y_q$ by introducing
$Y^0= |\det({\cal H})|_{W=0}$, the same determinant in the absence of disorder
(i.e. setting $W$ or $\tilde W$ to zero) and setting $E=0$.
Although some of our formula are valid for any $N$ our main results, i.e. the expression of the $\Sigma_q$, which we call the $q$ moment growth rates, will be obtained to the leading order at large $N$.

Note that the moments of the (absolute value of) spectral determinants (or, equivalently, of characteristic polynomials) of standard random matrices, Hermitian or unitary, as well as related objects have received considerable attention over the past 25 years, in great part due to
their role in developing conjectures for the behaviour of
the Riemann zeta-function on its critical line \cite{KS2000,BH2000,HKC01,my02,my03,AV03,BS06,FHK2012,FK2014}, see \cite{JKM23,BKrev} for a discussion and further references. The interest has been further boosted by relations to Fisher-Hartwig singularities of Toeplitz determinants, see e.g.\cite{Kra07,CLKra15,Fahs21,FyoLedouCounting2020}, and applications to Gaussian free fields, to freezing transitions \cite{FyoLeDouMoments2016} and to Gaussian multiplicative chaos, see e.g. \cite{Web15,BF22,LN24,SF25,Kel25}. More recently those studies have been extended to the non-Hermitian case \cite{WW19,Af19,L20,SSG23,SS25,Kivimae} and further to matrices with sparse or banded structure
\cite{Shch14,Shch17,AfSh25,Shch25}.

The moments of the spectral determinants of matrix valued random Schr\"odinger operators have so far received much less attention. In the scalar case $N=1$ and in $d=1$, the quantity $\Sigma_q$ turns out to be
identical to the so-called generalized Lyapunov exponent, called $\Lambda(q)$
 \cite{GLE,CPVbook}.
Such exponent describes the growth rate of the $q$-th power of the modulus of typical solutions of the initial value problem associated to random Schrodinger
equation \cite{TexierGLE2020} and was studied in the context of an elastic string
for any $q>-1$ in \cite{FLRT}.
The first moment, i.e. for $q=1$ was obtained for $N \to +\infty$ and for any $d$ in \cite{UsManifComplexity}, see also \cite{McKenna2}. The first moment is of special
interest in the context of disordered elastic systems
since it allows to compute the mean total number of stationary points ${\cal N}_{\rm tot} $
of the energy functional ${\sf H}[{\bf u}]$ via the application of the Kac-Rice formula (see \cite{RosmyRev,Fyo15} for an informal discussion in a general context, and
\cite{UsManifComplexity,McKenna2,FLRT} and references therein in the context of elastic manifolds):
\be \label{Nmoy}
\overline{{\cal N}_{\rm tot} }= \frac{ \overline{|\det {\cal K}[{\bf 0}]|}}{[\det(\mu - t \Delta)]^N} = \tilde Y_1.
\ee
The associated quantity $\Sigma_1$ has the interpretation of the annealed complexity of
such stationary points,
\be \label{defSigma}
\Sigma_1 = \lim_{L \to \infty} \frac{\log \overline{{\cal N}_{\rm tot} }}{N L^d}\,\,,
\ee
which was thus computed in the limit $N \to +\infty$ in \cite{UsManifComplexity} and
proved rigorously in \cite{McKenna2}.
\\

In this paper our goal is to
compute the more general quantities $\Sigma_q$. From these growth rates
we then obtain by Legendre transform the large deviation tail
of the distribution ${\cal P}({\sf e})$ of the
intensive "free-energy" defined as
\be
{\sf e} = \frac{1}{L^d N} ( \log |\det ({\cal H}- E)| - \log |\det ({\cal H})|  ).
\ee
Performing a saddle point in the large $L$ limit one obtains the large deviation result
\be
{\cal P}({\sf e}) \sim e^{ - N L^d \Phi({\sf e}) },
\ee
where the rate function $\Phi({\sf e})$ and $\Sigma_q$ are related by a Legendre transform
\be
\max_{{\sf e}} (q {\sf e }- \Phi({\sf e})) = \Sigma_q.
\ee
In $d=1$ and for arbitrary $N$, one can define $N$ Lyapunov exponents $\gamma_j$, $j=1,\dots,N$,
from the rate of growth of the volume spanned by $1\leq n \leq N$ independent solutions to the equation
$({\cal H}- E) \psi=0$.
It turns out that the quantity $N {\sf e}$ can be identified to the sum of these Lyapunov exponents, i.e.
$N {\sf e} = \sum_{j=1}^N \gamma_j$, see e.g. \cite{GrabschThesis},
and the rate function $\Phi({\sf e})$ thus describes
the fluctuations of this sum at large $N$, in a large deviation regime.
In the present context we define the associated generalised Lyapunov exponent by
\be \label{Lambdaq}
\Lambda(q,E)= \frac{1}{L} \log \langle e^{q L \sum_{j=1}^N \gamma_j } \rangle.
\ee

Before proceeding to the calculation, let us describe, following Ref. \cite{UsManifComplexity}, the main idea of one possible approach to
extracting the leading large $N$ asymptotics for such moments.
For clarity we consider the discrete model,
but the same applies to its continuum limit. Let us rewrite \eqref{Yq} for any $N$ as
\bea \label{main}
&& Y_q = \overline{  e^{  q \log |\det [ K + X + (\mu  - E)I ] | } } \\
&& = \prod_x \int_{\mathbb{R}}  \frac{d\xi(x) e^{- N \frac{\xi(x)^2}{2 c}}}{\sqrt{2 \pi/N}}
\langle e^{ q \log |\det [ K + X + (\mu  - E)I ] | }
\rangle_{\rm GOE's}\,,
\eea
where we used the decomposition \eqref{Hsum} for ${\cal H}$, $\langle \dots
\rangle_{\rm GOE's}$ denote averages over the i.i.d. GOE matrices $H(x)$
and we recall that $X_{ix,jy}=  J \xi(x) \delta_{ij} \delta_{xy}$.
Note that only the combination $\mu-E$ enters and below we will
use the notation freedom to use either parameter in different context.

In the large $N$ limit one can verify the important self-averaging property, to leading order in $N$
\be \label{selfaveraging}
\langle e^{ q \log |\det [ K + X + (\mu - E) I ] | }
\rangle_{\rm GOE's}
\approx
 e^{q \,\left\langle {\rm Tr}  \log | K + X + (\mu - E) I  |  \right\rangle_{\rm GOE's}}.
\ee
This property was conjectured in \cite{UsManifComplexity} and proved in \cite{McKenna1,McKenna2}
for $q=1$.
Using \eqref{selfaveraging} we arrive at
\be \label{start20}
Y_q|_{N \gg 1} \sim \prod_x \int_{\mathbb{R}}
 \frac{d\xi(x)}{\sqrt{2 \pi/N}}  e^{- N S[\xi] }
\ee
where the action $S[\xi]$ reads
\be \label{Sx}
S[\xi]= \sum_x \frac{1}{2 c} \xi(x)^2 - \frac{q}{N} \left\langle {\rm Tr}  (\log | K + X +  (\mu - E)I |
-  \log | - \Delta + \mu | )   \right\rangle_{\rm GOE's}
 \, .
\ee
The integral in \eqref{start20} is dominated at large $N$ by the saddle point for $\xi(x)$ which was studied for $q=1$
in \cite{UsManifComplexity,McKenna2} and will be studied here for general $q$. A crucial
property is that at the saddle point $\xi(x)$ is independent of $x$. This was proved using convexity arguments in
\cite{McKenna2}. Note that the non-triviality of the saddle point, and the non-linear dependence of $\Sigma_q$
on $q$ is induced by a non-vanishing $c>0$. On the other hand the spectral density of
the Schr\"odinger operator is independent on $c$ to leading order at large $N$ in the bulk
\cite{UsHessianManifold}.

In this paper we also develop a different and complementary method to study the moments $Y_q$, in the case $d=1$. In principle it applies to
any $N$ but we present here explicit results only in the large $N$ limit. As a byproduct it independently
verifies the self-averaging property \eqref{selfaveraging}.
Our method develops a connection \cite{OssipovYaglom}
between matrix valued Schrodinger operators and
a stochastic (non linear) matrix Ricatti equation. Passing to the corresponding
eigenvalues, and focusing on the continuum limit, leads to study a version of a Dyson Brownian motion (DBM)\cite{Dyson72} in a cubic (i.e. non confining) potential. The latter problem was
investigated in a different context by Allez and Dumaz \cite{AD15}.
They showed the existence of a phase transition in the limit $N \to +\infty$,
between a confined phase and a flowing phase for the DBM.
We use their results, and further develop the intriguing connections between the DBM in cubic potentials and matrix valued Schr\"odinger operators.
In particular we relate the rare activated barrier crossing events in the DBM
to the Lifschitz type tails of the density of states, and provide some large
deviation estimates at large but finite $N$. This method and these connections
are presented in Section \ref{sec:part1}.

In Section \ref{sec:part2} we develop the other method, inspired by Ref.
\cite{UsManifComplexity}, which was outlined above.
There we study the saddle point equation of the action \eqref{Sx} in any $d$,
and we compute $\Sigma_q$ as well as the large deviations rate function $\Phi({\sf e})$.
The particular cases of $d=0$ and $d=1$ are presented in more details.

\section{Continuum model in $d=1$: matrix Ricatti method} \label{sec:part1}

\subsection{General formalism}

In this section we consider the continuum model in $d=1$, and we set $\tilde \mu=0$ in \eqref{contmodel}, i.e.
we consider the $N \times N$ matrix operator with $\tau \in [0,L]$
\be
{\cal H}_{ij} = -  \tilde t \frac{d^2}{d \tau^2}  \delta_{ij} + \tilde W_{ij}(\tau)
\ee
where $\tilde t>0$. In the following we have set $\tilde t=1$.
Following closely the method developed by one of us in \cite{OssipovYaglom}, see Eqs (40-41) there, the functional determinant associated with this operator (with proper boundary conditions, here Dirichlet, see for $N=1$ Appendix A in Ref.
\cite{FLRT}, and properly regularized
\footnote{In the continuum limit a possible way of regularization is
to divide by the determinant of the free operator (obtained by setting $W=0$),
see e.g. \cite{FLRT,OssipovYaglom}.}) can be expressed as, see \ref{app:ricatti} :
\be \label{Trace}
\det({\cal H}-E) = y(L) \quad , \quad |y(L)| = e^{\int_0^L d\tau {\rm} {\rm Tr} Z(\tau) }
\ee
where the real symmetric $N \times N$ matrix $Z(\tau)$ satisfies the matrix Ricatti equation
\be \label{eqZ}
\partial_\tau Z = - E  - Z^2 + \tilde W(\tau)
\ee
with $Z(0)=+\infty$. This is the natural generalization to arbitrary $N$ \cite{OssipovYaglom} of the well
known $N=1$ Gelfand-Yaglom formula for the functional determinant of a 1D Schrodinger operator. In such a scalar case
it amounts to
solving the initial value problem for the function $y(\tau)$
\be
( - \tilde t \frac{d^2}{d \tau^2}  + \tilde W(\tau) - E ) y(\tau) =0 \quad y(0)=0 \quad , \quad y'(0)=1
\ee
which leads to $y(L)$ and the determinant in \eqref{Trace} at the final point $\tau=L$.
The latter method was used extensively in \cite{FLRT} to compute the annealed complexity of stationary points for an elastic string in a random potential.
\\

To analyze the matrix Ricatti equation \eqref{eqZ} we introduce the eigenvalues $\lambda_i(\tau)$ of the matrix
$Z$. Using standard perturbation theory in $W$ one can derive (see e.g. \cite{JPBBook,GautieThesis})
the stochastic evolution equation (in Ito convention)
\bea \label{Langevin1}
d \lambda_i(\tau) = - (E  + \lambda_i(\tau)^2 ) d\tau
+ \frac{\tilde J^2 d\tau }{N} \sum_{j \neq i} \frac{1}{\lambda_i(\tau) - \lambda_j(\tau)}
+ \frac{\sqrt{2 \tilde J^2}}{\sqrt{N}} dB_i(\tau)
+ \tilde J  \tilde \xi(\tau) d\tau
\eea
where the $dB_i(\tau)$, $i=1,\dots,N$ are independent standard Brownian motions.
For $E=0$, $c=0$ and in the absence of the quadratic term it is the standard equation for the
Dyson Brownian motion. It is important to note that here, because of the quadratic term,
the eigenvalues $\lambda_i(\tau)$ tend to blow up towards $-\infty$. When this happens they
are immediately reinjected at $+\infty$. This is a well known feature of the Ricatti method for for $N=1$
\cite{Halperin1965},
where $\lambda(\tau)=y'(\tau)/y(\tau)$ and this blow up corresponds to a zero of $y(\tau)$.
It generalizes to any $N$, as discussed in e..g \cite{AD15}, recalling that
the
initial condition is that all $\lambda_i(0)=+\infty$.
\\

To study these $N$ coupled stochastic equations we define the trace of the resolvent of the matrix $Z(\tau)$ as
\bea
G(z,\tau) = \frac{1}{N} \sum_i \frac{1}{\lambda_i(\tau)-z}
= \int \frac{d\lambda}{\lambda - z}  \rho(\lambda,\tau)
\eea
where $z$ has non zero imaginary part. $G(z,\tau)$ is thus the Stiljies transform of the (time-dependent) empirical density of eigenvalues
$\rho(\lambda,\tau) = \frac{1}{N} \sum_i \delta(\lambda_i(\tau)-\lambda)$. Standard methods, see \ref{app:resolvant}, allow to derive the exact stochastic evolution equation for $G(z,\tau)$:
\bea \label{eq5}
\partial_\tau G(z,\tau)  =  \frac{\tilde J^2}{2 N}   \partial_z^2 G(z,\tau)  +
   \partial_z [ z+ (E + z^2 -  \tilde J \tilde \xi(\tau)) G(z,\tau)]
+ \frac{1}{2} \tilde J^2  \partial_z G(z,\tau)^2 + \frac{\tilde J}{N} \partial_z \hat \eta(z,\tau)\,,
\eea
where $\hat \eta$ is a Gaussian noise with correlations
\bea \label{eq6}
\overline{\hat \eta(z,\tau) \hat \eta(z',\tau') } =  2 \frac{G(z,\tau)-G(z',\tau)}{z-z'} \delta(\tau-\tau').
\eea
The initial condition in the present problem, inherited from $\lambda_i(0)=+\infty$, is $G(z,0)=0$.
Until now this is exact for any $N$.

\subsection{Allez-Dumaz approach \cite{AD15} at $N=+\infty$}

Even in the large $N$ limit, \eqref{eq5} is non-trivial to solve for $c \neq 0$ because of the additional noise $\tilde \xi(\tau)$.
However, anticipating the use of the corresponding solution in the saddle point framework as explained
in the Introduction we can split this noise in two parts, a $\tau$-independent one (i.e. the zero mode, which is expected to be of order unity at the saddle point)
and a fluctuating one, of order $O(1/\sqrt{N})$ due to the Gaussian measure \eqref{Pxi} for $\xi(\tau)$, i.e. we decompose
\be
\tilde \xi(\tau) = \bar \xi + \frac{\eta(\tau)}{\sqrt{N}}.
\ee
Plugging into \eqref{eq5} and absorbing $\bar \xi$ in $E$ by redefining $E -  \tilde J \bar \xi \to E$
we obtain to leading order in the large $N$ limit
\bea \label{eq7}
\partial_\tau G(z,\tau)  =  \partial_z [ z+ (E + z^2 ) G(z,\tau)]
+ \frac{1}{2} \tilde J^2  \partial_z G(z,\tau)^2.
\eea

Remarkably, this problem appeared in quite unrelated context, in the work of Allez and Dumaz
\cite{AD15} about Hermitian matrix Brownian motion in a cubic potential.
The equation \eqref{Langevin1} setting $\xi(\tau)=\bar \xi$, redefining $E -  \tilde J \bar \xi \to E$
and taking $N \to +\infty$
describes the gradient dynamics of interacting particles
\be \label{Langevin2}
\frac{d}{d\tau} \lambda_i(\tau) = - V'(\lambda_i)
+ \frac{\tilde J^2  }{N} \sum_{j \neq i} \frac{1}{\lambda_i(\tau) - \lambda_j(\tau)}
\ee
in a cubic external potential
\be
V(\lambda) = E \lambda + \frac{\lambda^2}{3}.
\ee
Note that similar to the $N=1$ case the particles which go to $-\infty$ are immediately reinjected at $+\infty$.
For negative value $E<0$ the potential $V(\lambda)$ has a well of finite depth which, for
$N \to +\infty$, can trap the particles, as shown in \cite{AD15}.
As found there depending on the value of $E$ the steady state can be of two different types.
The correspondence of our parameters with \cite{AD15} is $a=-E$
and $\tilde J^2=\beta/2$ (strictly speaking $\beta=1$ for GOE, but in fact $\beta$
can be considered as a parameter in the analysis of \cite{AD15}). In the first phase, i.e. for
\be \label{Estar}
E < E^* =  -  \frac{3}{4} (2 \tilde J^2)^{2/3}
\ee
the particles are confined inside the well forming there a droplet, with the particle density having a finite support. In that phase there is no
current of particles going from $+\infty$ to $-\infty$. In the other phase, i.e. for $E>E^*$
the particles are pushed through the barrier by the inter-particle repulsion,
and the support of the density extends on the whole real axis.
In that phase there is a finite particle current.

The stationary solution of \eqref{eq5} with $E -  \tilde J \bar \xi \to E$ obeys
\be \label{statJ}
z+ (E + z^2) G(z) +  \frac{1}{2} \tilde J^2 G(z)^2 = {\cal J}\,,
\ee
where ${\cal J}= {\cal J}(E)$ is a (in general complex) integration constant to be determined. Given ${\cal J}$ there are
two solutions
\bea \label{Gpm}
G_\pm(z)= \frac{1}{{\tilde J}^2}  (- E - z^2 \pm \sqrt{(z^2 + E)^2 - 2 {\tilde J}^2 (z - {\cal J})})
\eea
and the problem is to find the value of the constant ${\cal J}$ and the branch which
leads to the proper solution, corresponding to a non-negative stationary density $\rho(\lambda)$.
This problem is solved in \cite{AD15}, where the constant ${\cal J}(E)$ is obtained in both phases,
as briefly recalled in the Appendix. In the confined droplet phase this constant is real, while in the
current phase it has an imaginary part. Indeed, as shown in \cite{AD15} the total particle current $j(x,t)$
goes to a constant $j$ in the steady state which is given by
\be \label{current}
j = \frac{1}{\pi}  {\rm Im} {\cal J}(E).
\ee

One can also relate the real part of ${\cal J}$ to the first moment of the stationary density of the Ricatti matrix eigenvalues. Indeed one has, on
one hand, by definition of the Stieljes transform, setting $z=\lambda + i 0^+$ and expanding for large $\lambda$:
\be
G(\lambda + i 0^+) = PV \int d\lambda' \frac{\rho(\lambda')}{\lambda'-\lambda} + i \pi \rho(\lambda)
\simeq - \frac{1}{\lambda} - \frac{1}{\lambda^2} \int d\lambda' \lambda' \rho(\lambda') + O(\lambda^{-3})
+ i \pi \rho(\lambda).
\ee
On the other hand the expansion of \eqref{statJ} at large $z$ gives
\be
G(z) = - \frac{1}{z} + \frac{{\cal J}}{z^2} + O(\frac{1}{z^3}).
\ee
Identifying one obtains
\be \label{limitTrace}
\lim_{N \to +\infty}  \left\langle \frac{1}{N}  {\rm Tr Z} \right\rangle_{st}  = \langle \lambda \rangle_{st} = \int d\lambda \lambda \rho(\lambda) = - {\rm Re} {\cal J}(E)
\ee
Note that this also implies that the density decays at large $\lambda$ as
\be
\rho(\lambda) \simeq \frac{1}{\pi \lambda^2} {\rm Im} {\cal J} + O(\lambda^{-3})\,,
\ee
with ${\rm Im} {\cal J}  >0$ only in the current phase.

Recalling that $\det({\cal H} - E)=y(L)$ and using  \eqref{limitTrace}
allows us after restoring $E \to E-  \tilde J  \bar \xi$ in the r.h.s. of \eqref{limitTrace}, to rewrite Eq. \eqref{Trace} as
\be
\lim_{N \to +\infty}   \frac{1}{N L} \log |y(L)| = \lim_{N \to +\infty}  \frac{1}{N L}  \int_0^L dx {\rm} Tr Z(x) =
 \frac{1}{L}  \int_0^L dx \int d\lambda \lambda \rho(x,\lambda)
 \to_{L \to +\infty} \langle \lambda \rangle_{st} =  - {\rm Re} {\cal J}(E-  \tilde J  \bar \xi)
\ee
where we have assumed that in the limit of large $L$ the
integral is dominated by the stationary solution.

Further recalling the decomposition ${\cal H}= K + X + \mu I$, and identifying in the present
continuum model $K= - \tilde t \partial_\tau^2 I + \tilde J \tilde H(\tau)$, $X= \tilde J \tilde \xi$,
and $\mu=0$,  one finds that
\be  \label{relation2}
- \frac{d}{d E} {\rm Re} {\cal J}(E -  \tilde J \bar \xi) =  \lim_{L,N \to +\infty}  \frac{1}{N L} {\rm Re}
{\rm Tr} \frac{1}{ E - {\cal H}  + i 0^+ }
= - PV \int d\alpha \frac{\rho_K(\alpha) }{\alpha - E + \tilde J  \bar \xi}\,,
\ee
where we denote $\rho_K(\alpha)$ the mean
spectral density of $K$, which is known, see next Section.
\\

Using the above results we now can evaluate, with averaging over the uniform zero mode $\bar \xi$:
\be
Y_q = \overline{ |\det( {\cal H}-E) |^q }  =
\overline{ |y(L)|^q } \sim  \int d\bar \xi e^{- N L \frac{ \bar \xi^2 }{2 c} } e^{ - q N L {\rm Re} {\cal J}(E-  \tilde J \bar \xi)  }.
\ee
Evaluating this integral via saddle point we find that the saddle point is at $\bar \xi = \xi^*_q$, which using \eqref{relation2} is given by
\be
\frac{\xi^*_q}{ c \tilde J} = q PV \int d\alpha \frac{\rho_K(\alpha) }{\alpha - E +  \tilde J \xi^*_q}.
\ee
In the next Section we will see that the same action and saddle point equation arise
in the alternative method summarized in the introduction, showing equivalence
between the two approaches. This equivalence is not {\it a priori} trivial, as the two methods employ different order of limits: $\lim_{N\to \infty}\lim_{L\to \infty}$ in the DBM/Ricatti approach vs. $\lim_{L\to \infty}\lim_{N\to \infty}$
in the saddle-point method. It is well-known that properties of $1d$ matrix-valued Schr\"odinger Hamiltonians (or their discrete analogues) such as
local eigenvalue and eigenfunction statistics are very sensitive to the order of limits. Namely,  taking the limit $N \to \infty$ first ensures Wigner-Dyson eigenvalue statistics in the bulk \cite{TShch14} accompanied by full eigenfunction delocalization \cite{Stoneetal25}, while sending the system length $L\to \infty$ first ensures complete localization and Poisson statistics. In contrast, the large deviation function controlling the spectral determinant growth
in the spectral bulk  turns out to be insensitive to the order of limits.
This result emphasizes that the fluctuations of $\log \det({\cal H}-E)$ in the large $N$ limit,
reflected in the large deviation function, are controlled to the leading order solely
by the fluctuations of the zero mode
$\bar \xi$ (which exist only for $c>0$).
\\

\subsection{Connection to our model for $N \gg 1$, density of states and barrier crossing}
\label{sub:connection}

In this Section we describe in more details the connection between the DBM
in the cubic potential studied in the Allez-Dumaz paper \cite{AD15} and the properties of
matrix valued random Schrodinger operator with $c=0$.

In the DBM, $E=-a$ is a parameter which controls the depth of the cubic
potential well and for $N=+\infty$ a transition occurs at $E=E^*$. If the potential well is
deep enough, i.e. $E<E^*$, the DBM particles are confined by a barrier, while
if $E>E^*$, the DBM particles can flow to $\lambda=-\infty$ and be reinjected back at $\lambda=+\infty$. We first show that the
critical value $E^*$ corresponds to the edge of the spectrum of the
random Schrodinger operator in the limit $N \to +\infty$, with a non zero density of states (DOS) for $E>E^*$,
and a vanishing DOS for $E<E^*$.

The spectral density $\rho_K(\alpha)$ of the operator $K= - \tilde t \partial_\tau^2 I + \tilde J \tilde H(\tau)$ in $d=1$ was computed
in the large $N$ limit taken first by two of us in \cite{UsHessianManifold}, as will be recalled below in the more
general context of arbitrary $d$ and discrete models. Note that it does not depend on $c$.
We have checked that
our result for that density can be related to the Allez-Dumaz calculation
as follows
\be \label{conjecture}
\rho_K(\alpha) = \frac{1}{\pi}  \frac{d}{dE} {\rm Im} {\cal J}(E)|_{E=\alpha}
\ee
where the explicit formula involves the roots of a cubic equation
and is given in the \ref{app:cubic}.

The relation \eqref{conjecture} can be understood as a generalization to large $N$ of the relation between
the integrated density of states and the current of particles $j$ in the Ricatti variable
(crossing from minus to plus infinity, i.e. the number of explosions)
which was discovered for $N=1$ in \cite{Halperin1965}. From \eqref{current}
this leads to \eqref{conjecture}. Indeed the relation between the eigenvalue
counting of a stochastic Schrodinger operator and the rate of
explosions of its associated Riccati equation (so called oscillation
theorems) have been extended to any $N>1$ for a class of matrix valued
random operators  \cite{BaurKratz,Kratz} (for recent applications
see e.g. \cite{RiderValkoBlock})

Coming to the case of large but finite $N$ it is natural to expect that
activated barrier crossing for the DBM will occur leading to a small
but non zero current. In the framework of random matrix valued Schrodinger
operator this corresponds
to the fact that at finite $N$ the tail of the DOS extends to the region $E<E^*$.
For $N=1$ this is the famous Lifschits tail regime with a density
$\sim \exp(- c |E|^{3/2})$ for large negative $E$. For large $N$
it is expected to be also exponentially suppressed in $N$.

To estimate these tails in the DOS in the region $E<E^*$ for large $N$ one may relate it to computing the
barrier crossing probability for the DBM  in the confined phase.
This is done in \ref{app:cubic}.
 It is natural to assume that the barrier crossing is dominated by "single particle processes"
since collective jumps over the barrier are presumably much less likely.
We thus consider the probability of a single (leftmost) eigenvalue of the
associated Ricatti matrix to leak over the barrier towards $-\infty$.
The crossing time is given by the Arrhenius formula
\bea
\tau_{\rm crossing} \sim \exp(  \frac{N U}{\tilde J^2} )
\eea
where $U$ is the effective barrier which is explicitly computed in \eqref{barrierexact}.
 We have used that the Brownian noise in \eqref{Langevin1} corresponds
to an effective "temperature" $\tilde J^2/N$. Since the number of barrier crossing
corresponds to the counting of the energy levels, this leads to the following asymptotics for the average DOS
outside of the spectrum:
\bea
\rho_K(\alpha) \sim \tau_{\rm crossing}^{-1} \sim \exp( -  \frac{N U}{\tilde J^2} ).
\eea
Close to the edge of the spectrum (i.e. close to the
transition at $E=E^*$ and for $E<E^*$) the barrier behaves as
\be
U \simeq \frac{4}{5} \sqrt{2} \times 3^{1/4} \beta^{1/6}  \,  (E^*-E)^{5/4}
\ee
while in the limit $E \to - \infty$ the behavior is
\be
U \simeq  \frac{4}{3} |E|^{3/2}.
\ee
 The $3/2$ exponent at large $E$ can be obtained simply from
the original barrier of the cubic potential. However
the appearance of an exponent $5/4$ is a novel feature of the present problem.
Near criticality (i.e. near the edge of the spectrum) the barrier is strongly renormalized by the
log-interaction in the DBM and the new exponent arises from the unusual form form the droplet density
at criticality (which vanishes at the edge with a $3/2$ exponent, different from the usual semi-circle density
exponent $1/2$).

We now turn to the analysis of the solutions of the saddle point equation in the more general
context of the $d$ dimensional model and $c>0$.

\section{Discrete and continuum model, any $d$: extension of previous method} \label{sec:part2}

\subsection{Saddle point equations and determination of $\Sigma_q$}

We now go back to the discrete model \eqref{discretemodel} and compute the reduced moments $\tilde Y_q$
defined in \eqref{Yq}, i.e.
$\tilde Y_q= \overline{
( \det |{\cal H}| /|\det|{\cal H}||_{W=0})^q}$
where ${\cal H}=K + X +  \mu I$ is defined in \eqref{Hsum}.
Here we have set the variable $E=0$. Below when treating the $d=1$
case we will reinstate $E$. The results can be expressed in full generality only in terms of the eigenvalues of the operator
$\Delta_{xy}$, which we denote $\Delta(k)$. For concreteness, and for more explicit
calculations, we will choose $\Delta_{xy}$ to be the discrete Laplacian with periodic boundary conditions.
In that case the Laplacian eigenmodes are plane waves $\sim e^{i k \cdot x}$ and the associated eigenvalues
are denoted by $\Delta(k)$. For instance in $d=1$ one has $\Delta(k)= 2 (\cos k-1)$
with $k=2 \pi n/L$, $n=0,..L-1$.  One can obtain the continuum model from the discrete one in $d=1$,
setting $L=M a$ and taking $M \to +\infty$, $a \to 0$ at fixed $L$. One also sets $\tau = x a$ and $\tilde t=t/a^2$,
and the continuum Fourier variable is $\tilde k$ with $k=a \tilde k$
so that as $a \to 0$ one has $- t \Delta(k) = 2 t (1-\cos k) \to  \tilde k^2$.
As $a \to 0$ the discrete operator converges to the continuum one in the sense that
$g^i_x = \sum_y {\cal K}_{ix,jy} g^j_y \quad \Leftrightarrow g^i(\tau) = {\cal H}_{ij} . f^j(\tau)$
with the correspondence $f_y=f(y a)$ and $g_x=g(x a)$. The replacement $- t \Delta(k)  \to  \tilde k^2$ can be similarly made to obtain the continuum model in any $d$. Note that in all cases
the operator in the absence of disorder, noted ${\cal H}^0$, has a positive spectrum,
i.e. $\mu - t \Delta(k) \geq 0$.
\\

Let us start from \eqref{start20} and \eqref{Sx} which we recall here
\be \label{start2}
\tilde Y_q|_{N \gg 1} \sim \prod_x \int_{\mathbb{R}}
 \frac{d\xi(x)}{\sqrt{2 \pi/N}}  e^{- N S[\xi] } \quad , \quad
 S[\xi]= \sum_x \frac{1}{2} \xi(x)^2 - \frac{q}{N}
 \left\langle {\rm Tr}  (\log | K + X +  \mu I |
-  \log | - \Delta + \mu | )   \right\rangle_{\rm GOE's}
 \ee
where $X_{ix,jy}=  J \xi(x) \delta_{ij} \delta_{xy}$, $K$ is defined in \eqref{Hsum}
{and ${\rm Tr}$ denotes the trace over the $MN$ dimensional space.
We have set $c=1$ for simplicity. Note that the regularized quantity $\tilde Y_q$
remains finite in the continuum limit.
To obtain the asymptotic of this integral at large $N$ we will use the saddle point method and
look for the minimum of the action $S[\xi]$. This optimal configuration must satisfy
\be
\xi(x) = \frac{q}{N} J \left\langle {\rm tr}  (K + X +  \mu I)^{-1}_{xx}  \right\rangle_{\rm GOE's}
\ee
where here ${\rm tr}$ denotes the trace only over the $N$ dimensional space.
This equation has at least one solution independent of $x$, $\xi(x)=\xi^*_q$. In the case $q=1$ it was
proved to be unique, and to correspond to the absolute minimum of $S[\xi]$ \cite{McKenna2}.
We
proceed assuming that this property still holds. The parameter $\xi^*_q$ is the solution of the
equation
\be \label{sp0}
\xi^*_q = q f'(\xi^*_q + \frac{\mu}{J})  \quad , \quad  f(\xi) := \int d\alpha \ln|\alpha+ J \xi| \, \rho_{K}(\alpha)
\ee
where $\rho_K(\alpha)$ is the mean eigenvalue density of the random matrix $K$
in the limit $N \to +\infty$, which was determined by two of us in \cite{UsHessToy}.
Using in this Section the notations of the companion paper
 \cite{UsManifComplexity},
it is given by the imaginary part of the resolvent $i r_\lambda$ defined below
\be \label{dens}
\rho_{K}(\lambda) = \frac{1}{\pi} {\rm Im} (i r_\lambda)|_{{\rm Im} \lambda=0^-}~,~
i r_\lambda := \frac{1}{N M} \langle {\rm Tr} (\lambda - K)^{-1} \rangle
= \int d\alpha \frac{\rho_{K}(\alpha)}{\lambda-\alpha}\,,
\ee
which satisfies the following self-consistency Pastur-type equation \cite{UsHessToy,KhorPast1993}
\be \label{sc}
i r_\lambda = \int_k \frac{1}{\lambda +  t \Delta(k)   - i r_\lambda J^2}\,,
\ee
where we denoted
\be
\int_k = \frac{1}{M} \sum_k  \equiv \int \frac{d^d k}{(2 \pi)^d}
\ee
making our formulas valid both for discrete and continuum models (in the latter case $\sum_x \equiv \int d^dx$).
The growth rate of the $q$-th reduced moment $\Sigma_q$ defined in \eqref{defSigmaq}
is obtained from the value of the action $S[\xi]$
in \eqref{start2} at the saddle point as \footnote{In calculations below we
should keep in mind that in the continuum limit and for $d>0$
the integral in \eqref{sp0} which defines the function $f(\xi)$ diverges at large $\alpha$.
However all our results depend only on the combination $f(\xi + \frac{\mu}{J}) -  \int_k  \ln (\mu -  t \Delta(k))$
which is finite. }
\be \label{sig1}
\Sigma_q =  - S(\xi_q^*) - q \int_k  \ln (\mu -  t \Delta(k))=  - \frac{1}{2} (\xi_q^*)^2 +  q f(\xi_q^*+ \frac{\mu}{J})  - q \int_k  \ln (\mu -  t \Delta(k)),
\ee
where $\xi_q^*$ is obtained by solving \eqref{sp0}.

Let us obtain an equivalent but more convenient set of equations to determine $\xi_q^*$
and $\Sigma_q$. To this aim we first note that at the saddle point for $\xi$, we have
\bea \label{sp2}
\xi_q^* = - q J {\rm Re} [ i r_\lambda] |_{\lambda=- J \xi_q^*-  \mu + i 0^+}
\eea
since by definition of $f(\xi)$ one has
\be
f'(\xi) = J \, PV \int  \frac{d\alpha \rho_{K}(\alpha)}{J \xi+\alpha}
= J \, {\rm Re}\int \frac{d\alpha \rho_{K}(\alpha)}{J \xi+\alpha- i 0^+}
=
- J \, {\rm Re} ( i r_{- J \xi + i 0^+} ), \nonumber
\ee
where the last equality follows from the definition \eqref{dens}.
To evaluate the real part in the r.h.s. of \eqref{sp2}
we separate the real and imaginary parts as
\bea
i r_\lambda = x_\lambda + i y_\lambda.
\eea
The equation \eqref{sc} then leads to the equivalent pair of equations
\bea \label{solu2}
&& x_\lambda = \int_k \frac{ \lambda - J^2 x_\lambda
+   t \Delta(k)}{(\lambda - J^2 x_\lambda +   t \Delta(k))^2 + J^4 y_\lambda^2},  \\
&& y_\lambda = y_\lambda \int_k \frac{J^2}{(\lambda -  J^2 x_\lambda +   t \Delta(k))^2 + J^4 y_\lambda^2},
\eea
where $y_\lambda \geq 0$.
Substituting $\lambda=- J \xi^*_q-  \mu$ in these equations, one obtains
the set of equations
\bea \label{solu0}
&& \xi^*_q = J q \int_k \frac{ \mu_q -   t \Delta(k)}{( \mu_q -  t \Delta(k))^2 + J^4 y^2},  \\
&& y = \int_k \frac{y J^2}{( \mu_q -   t \Delta(k))^2 +  J^4 y^2},  \label{soluy} \\
&& \mu_q = \mu + (1- \frac{1}{q})J \xi^*_q,
\eea
since $\lambda- J^2 x_\lambda|_{\lambda=- J \xi^*_q -  \mu}=-\mu_q$
using \eqref{sp2}. Here $y \geq 0$ is a variable which
should be eliminated between the above equations to obtain the value $\xi_q^*$
at the saddle point. Once it is obtained it can be inserted into
\eqref{sig1} to obtain $\Sigma_q$.

As we will discuss in more detail below, there are two
phases depending on whether $y=0$ or $y >0$.
Noting, from \eqref{dens}, that
\be \label{remark}
y=y_\lambda|_{\lambda=- J \xi^*_q-\mu}= \pi \rho_K(-  J \xi^*_q-\mu),
\ee
we see that these two phases also correspond to $- J \xi^*_q-\mu$
belonging or not to the support of the mean eigenvalue density of $K$.\\

Let us give two useful identities which are consequence of the saddle point equation
and which allow to compute how $\Sigma_q$ varies when either $\mu$ or $q$
are varied. First,
taking a derivative w.r.t. $\mu$ in \eqref{sig1} and
using the saddle point condition \eqref{sp0} one obtains for any $q$
\bea \label{dersigmu}
&& \partial_\mu \Sigma_q =
\frac{q}{J} f'(\xi^*_q + \frac{\mu}{ J}) - q \int_k \frac{1}{\mu -   t \Delta(k)}  =\frac{1}{J} \xi^*_q  - q \int_k \frac{1}{\mu -  t \Delta(k)}.
\eea
Doing the same by instead taking a derivative w.r.t. $q$ in \eqref{sig1} and
using the saddle point condition \eqref{sp0} one obtains for any $q$
\bea \label{dersigq}
&& \partial_q \Sigma_q = f(\xi^*_q + \frac{\mu}{ J}) - \int_k \log(\mu -   t \Delta(k)).
\eea

\subsection{Legendre transform and the rate function for ${\sf e}$} We now compute the large deviations (LD) of the random variable
${\sf e} =  \log \frac{\det |{\cal H}|}{\det |{\cal H} ^0|}$ at large $N$.
The general formula for the LD rate function is given by
\be
\Phi({\sf e}) = {\rm max}_q [ q \, {\sf e}  - \Sigma_q ].
\ee
Generically, the derivative conditions 
\be
\Phi'({\sf e}) = q \quad , \quad \partial_q \Sigma_q = {\sf e}
\ee
give a relation between the optimal $q$ and ${\sf e}$. The
second equation, using \eqref{dersigq}, gives the relation
\be \label{elegendre}
{\sf e}  = \partial_q \Sigma_q = f(\xi_q^* + \frac{\mu}{J}) -  \int_k \log(\mu -   t \Delta(k)).
\ee
The typical, i.e. the most probable, value ${\sf e}_{\rm typ}$ of ${\bf e}$ is defined by $\Phi'({\sf e}_{\rm typ})=0$. Hence it corresponds to $q=0$. Evaluating \eqref{elegendre} at $q=0$ we obtain
\be \label{etyp}
{\sf e}_{\rm typ} = \partial_q \Sigma_q|_{q=0} = f(\frac{\mu}{J}) -  \int_k \log(\mu -   t \Delta(k))\,,
\ee
since for $q=0$ one has $\xi_q^*=0$, as can be seen from
\eqref{sp0}. Note that using the definition of $f(\xi)$ in \eqref{sp0}
one can rewrite \eqref{etyp} as
\be
{\sf e}_{\rm typ}
= \overline{ \frac{1}{N} {\rm Tr} \log |K + \mu I | } - \frac{1}{N} {\rm Tr} \log(\mu I -   t \Delta(k))
= \overline{ \frac{1}{N} {\rm Tr} \log |{\cal H}| } - \frac{1}{N} {\rm Tr} \log |{\cal H}^0|\,,
\ee
which shows that the noise $\xi$ is irrelevant for computing the mean resolvent, i.e. for typical
values of ${\bf e}$, as noted in \cite{UsHessToy,UsHessianManifold}.
\\

To study the LD rate function $\Phi({\sf e})$, it is then natural to introduce the difference
\be
\delta {\sf e} = {\sf e}- {\sf e}_{\rm typ}
\ee
which, upon substracting \eqref{etyp} from \eqref{elegendre} gives
\be
\delta {\sf e}  = f(\xi_q^* + \frac{\mu}{J}) - f(\frac{\mu}{J}).
\ee
Introducing $f^{-1}(a)=\xi$ the inverse function of $f(\xi)=a$ (assuming that $f(\xi)$ is
monotonically decreasing) one obtains
\be
\xi_q^* = f^{-1} \left(  \delta {\sf e}  + f(\frac{\mu}{J}) \right)  - \frac{\mu}{J}
\ee
On the other hand one can also rewrite, using \eqref{elegendre}
\be
\Sigma_q =   - \frac{1}{2} (\xi_q^*)^2 +  q f(\xi_q^* + \frac{\mu}{J})  -  q \int_k \log(\mu -   t \Delta(k))\,,
=  - \frac{1}{2} (\xi_q^*)^2 +  q {\sf e}
\ee
which leads to our main general result for the rate function $\Phi({\sf e})$
\be \label{resPhi}
\Phi({\sf e}) = q {\sf e}  - \Sigma_q = \frac{1}{2} (\xi_q^*)^2 = \frac{1}{2} \left( f^{-1} \left( \delta {\sf e}  + f(\frac{\mu}{J}) \right) - \frac{\mu}{J}  \right)^2.
\ee
It is particularly convenient since one notes that the function $f(z)$ can be shifted by any arbitrary constant
$f(z) \to f(z) + b$, without changing the result. We will make use of this result below in specific examples.
Note that expanding \eqref{resPhi} in series of $\delta {\sf e}$ we obtain that the distribution of ${\sf e}$ is Gaussian near its typical value, and determine its variance
\be
\Phi({\sf e}) = \frac{(\delta {\sf e})^2}{2 f'(\frac{\mu}{J})^2} -
\frac{f''(\frac{\mu}{J}) (\delta {\sf e})^3}{2 f'(\frac{\mu}{J})^4} +
O((\delta {\sf e})^4) \quad , \quad
{\rm Var} \, \delta {\sf e} =  f'(\frac{\mu}{J})^2. \label{vare}
\ee
\\

\subsection{Phase diagram}

We now discuss the solutions of the above saddle point equations. In the plane $q,\mu$ there are two phases.
For $q=1$ these correspond to (i) the simple phase of the associated elastic manifold problem, with zero complexity $\Sigma_1=0$ and (iii) the complex phase with $\Sigma_1>0$. The phase transition corresponds to landscape
topological trivialization \cite{UsTrivialization}.
Although there is no such interpretation
for $q \neq 1$ we will retain this terminology for convenience. In particular
we find that the variance of ${\sf e}$ is non-analytic at this transition, see below.
\\

{\bf Simple phase}. In this phase $y=0$ at the saddle point. One has from \eqref{solu0} that $\xi_q^*$ is then the
solution of
\be \label{solu2}
 \xi^*_q = J q \int_k \frac{1}{\mu_q -   t \Delta(k)} \quad , \quad \mu_q = \mu + (1- \frac{1}{q})J \xi^*_q.
\ee
For $q=1$ one has $\mu_q=\mu$ and the equation simplifies with
$\xi^*_1 = J  \int_k \frac{1}{\mu -   t \Delta(k)}$. Let us recall why
in that phase the complexity vanishes, $\Sigma_1=0$. Indeed, consider
the equation \eqref{dersigmu} for $\partial_\mu \Sigma_q$. The right hand side vanishes,
hence this derivative vanishes for $q=1$. Since $\Sigma_1|_{\mu =+\infty}=0$
we obtain that $\Sigma_1$ is zero everywhere in this phase. This however is
true only for $q=1$, and for general $q$ the solution is more complicated.
It will be studied on some examples below.
\\

{\bf Complex phase}. In this phase $y > 0$. One can then simplify \eqref{soluy} and obtain
\be \label{solu3}
1 = J^2 \int_k \frac{1}{( \mu_q -   t \Delta(k))^2 +  J^4 y^2}  \quad , \quad \mu_q = \mu + (1- \frac{1}{q})J \xi^*_q
\ee
which together with \eqref{solu0} which we recall here
\be \label{solu4}
 \xi^*_q = J q \int_k \frac{ \mu_q -   t \Delta(k)}{( \mu_q -  t \Delta(k))^2 + J^4 y^2}  \\
\ee
allows to determine $y$ and $\xi^*_q$. Whenever the transition to the simple phase is continuous, one can obtain the boundary of this phase by letting $y \to 0^+$,
which leads to
\bea  \label{boundary0}
&& 1 = J^2 \int_k \frac{1}{( \mu_q -   t \Delta(k))^2}  \quad , \quad \mu_q = \mu + (1- \frac{1}{q})J \xi^*_q, \\
&& \xi^*_q = J q \int_k \frac{ 1}{ \mu_q -  t \Delta(k)}.
\eea
Therefore the phase boundary
is given by $\mu_q= \mu_c$ where
$\mu_c$ is the so-called Larkin mass, which is the unique solution of
\be \label{Larkin}
1 = J^2  \int_k \frac{1}{( \mu_c -  t \Delta(k))^2}.
\ee
This equation is the so-called replicon instability condition, which signal a continuous transition towards a replica symmetry breaking phase for $\mu < \mu_c$ in the corresponding
statistical mechanics model at $T=0$ \cite{MezPar91}. From \eqref{boundary0} we finally
get that the transition to the complex phase occurs as $\mu=\mu_b$ given for general $q$ as
\be \label{mubgen}
\mu_b = \mu_c - (1 - \frac{1}{q}) J \xi^*_q = \mu_c - (q - 1) J^2
 \int_k \frac{1}{\mu_c -  t \Delta(k)}
\ee
Below we analyze a few cases where these equations can be solved explicitly.
\\

\subsection{Explicit solution for $d=0$}.
Consider $d=0$, i.e. the Hessian problem of a single particule in a random potential,
discarding the elastic energy given
by the Laplacian term. In such a case ${\cal H}=W+ \mu I=K+X+\mu I$ is simply a random matrix with correlations
as in \eqref{correlatorW} with $c=1$.
The spectrum of $K$ is then a semi-circle with
$\rho_K(\alpha) = \frac{1}{2 \pi J^2} \sqrt{4 J^2 - \alpha^2}$.
One thus needs to find the minimum of the action $S(\xi)= \frac{1}{2} \xi^2 - q f(\xi + \frac{\mu}{J})$
where, from the definition \eqref{sp0}
\bea  \label{fzerod}
&& f(\xi) = \log J + \frac{\xi^2}{4} - \frac{1}{2} \quad , \quad |\xi| < 2. \\
&& f(\xi) = \log J + \frac{1}{4} (\xi^2-2 - |\xi |\sqrt{\xi^2-4} + 4 \log( |\xi | + \sqrt{\xi^2-4}) - 4 \log 2 \quad , \quad |\xi | > 2.
\eea
Note that this function is not analytic at $\xi=2$ since, expanding around this point, one has
\bea
&& f(\xi) \simeq  \log J +  \frac{1}{2}+(\xi-2)+\frac{1}{4}
   (\xi-2)^2   \quad , \quad \xi<2. \\
   && f(\xi) \simeq   \log J + \frac{1}{2}+(\xi-2)-\frac{2}{3} (\xi-2)^{3/2}+\frac{1}{4}
   (\xi-2)^2 \quad , \quad \xi>2.
\eea
so that the first and second derivatives are continuous at $\xi=2$, but the second derivative is singular.
Indeed, one has
\be
f''(\xi)= \frac{1}{2} -  \theta(|\xi|>2) \frac{1}{2} \frac{|\xi|}{\sqrt{\xi^2-4}}.
\ee
Hence
\be
S''(\xi) = \begin{cases} \frac{1}{2} (1-q) \quad , \quad |\xi + \frac{\mu}{J}| < 2 \\
\frac{1}{2} + \frac{q}{2} (  \frac{|\xi|}{\sqrt{\xi^2-4}}  - 1) \quad , \quad |\xi + \frac{\mu}{J}| > 2
\end{cases}
\ee
so that for $0<q<1$ the action is convex and a unique minimum is guaranteed.
\\

On the other hand, Eq. \eqref{Larkin} gives the Larkin mass $\mu_c=J$ in $d=0$, which in turns gives the boundary of the
two phases as $\mu_b = (2-q) J$.
\\

{\bf Simple phase}. In the simple phase, $\mu>\mu_b=(2-q)J $, from \eqref{solu2} one finds
 the following equation for $\xi_q^*$
\be
\xi^*_q = \frac{J q}{\mu + (1 - \frac{1}{q}) J \xi^*_q}
\ee
leading to
\be \label{xizerod}
\xi^*_q = \frac{2 J q}{\mu + \sqrt{\mu^2 + 4 J^2 (q-1)} }.
\ee
Note there are actually two branches. However the proper solution is continuous at $q=1$ hence
corresponds to the $+$ branch. Note also that $\mu \geq  \mu_b=(2-q)J$ implies that
$\mu^2 + 4 J^2 (q-1)  \geq 0$.
Hence this branch is the correct one (as can also be checked by comparing the
actions at the saddle point as shown below).
\\

To obtain the growth rate $\Sigma_q$, defined in \eqref{defSigmaq}, of the moments
$\tilde Y_q= \overline{ \det|{\cal H}|^q/\mu^{q N}}$
in the simple phase for general $q$, i.e. for $\mu > \mu_b = (2- q)J $,
we first note that this condition 
implies that
$\xi^*_q  + \frac{\mu}{J} > 2$ for $\xi^*_q$ given by \eqref{xizerod},
hence to evaluate \eqref{sig1} we must use the second line in \eqref{fzerod}.
After some simplifications one finds $\Sigma_q$ for $\mu>(2-q)J $ as
\be
 \Sigma_q = -\frac{q \left(\mu ^2-\mu  \sqrt{\mu ^2+ 4 J^2(
q-1)}\right)}{4 J^2 (q-1)} + q \log \left(1+\sqrt{1+ 4 \frac{J^2}{\mu^2}( q-1) } \right) - \frac{q}{2} (1+2 \log 2 )  \quad , \quad \mu \geq (2-q) J
\ee
The behavior near $q=0$ allows to recover the mean \eqref{etyp} and the variance of ${\sf e}$ \eqref{vare}
\be
{\rm Var} \, {\sf e}= \frac{1}{2 J^2} \left(\mu^2- \mu \sqrt{\mu ^2-4 J^2}  -2 J^2 \right) = f'(\mu/J)^2.
\ee
One checks that $\Sigma_q$ vanishes as $q \to 1$ in the simple phase as
\bea
 \Sigma_q= \frac{(q-1) J^2}{2 \mu ^2}+\frac{\left(\mu ^2-J^2\right) J^2
   (q-1)^2}{2 \mu ^4}+O\left((q-1)^3\right).
\eea

If $q > 2$ there is only a simple phase for any $\mu > 0$,
and the minimum of the action is given by \eqref{xizerod}. Note that
for $q > 2$ and $\mu=0$ the action $S(\xi)$ has a symmetric double well form
and there are then two equivalent minima at $\xi = \pm q/\sqrt{q-1}$.
Any small $\mu>0$ breaks the degeneracy and leads to the $+$ branch in \eqref{xizerod}.
Hence for $q>2$ the simple phase extends to the value $\mu=0^+$ and the rate
$\Sigma_q \sim - 2 q \log \mu$ diverges logarithmically there.

If $0<q <2$ there is a transition at a positive
value of $\mu$, and the value at the transition is
\be
\Sigma_q |_{\mu=\mu_b=2-q} =-\frac{1}{2} q (q+2 \log (2-q)-1)
\ee
which is negative for $q<1$ and positive for $q>1$.
\\

{\bf Complex phase}. In the complex phase in $d=0$, $\mu<\mu_b=(2-q) J$
the solution of \eqref{solu3}, \eqref{solu4} is given by
\be
\mu_q = \frac{\mu}{2 - q} \quad , \quad \xi_q^* = \frac{q}{2-q} \frac{\mu}{J} \quad , \quad y^2 = \frac{1}{J^2} (1 - \frac{1}{(2-q)^2} \frac{\mu^2}{J^2}).
\ee
One checks that $\xi^*_q  + \frac{\mu}{J} = \frac{2}{2-q} \frac{\mu}{J} = 2 \frac{\mu}{\mu_b} <2$, hence to evaluate \eqref{sig1} we must use the first line in \eqref{fzerod} and we obtain
\be \label{complexSigmad0}
\Sigma_q = \frac{q}{2 (2-q) } \frac{\mu^2}{J^2} + q \log \frac{J}{\mu} - \frac{q}{2} \quad , \quad 0 \leq \mu \leq (2-q) J.
\ee
Note that the branch $|\xi|<2$ of $f(\xi)$ in the first line in \eqref{fzerod} (which corresponds to
the complex phase) leads to $S''(\xi) =\frac{1}{4}(2- q)$ so a minimum can exist only for
$q<2$, consistent with the above result (in other words, the above saddle point in complex phase becomes
a local maximum of $S(\xi)$ for $q>2$)
\\

For $q=2$ and $\mu>0$ the minimum is at $\xi_q^*$ given by the simple phase result \eqref{xizerod}.
Note the peculiarity that as $\mu \to 0$ and $q=2$ the function $S(\xi)$ is exactly constant in the region
$|\xi|<2$ hence reaches its minimum everywhere on this interval.
\\

{\bf Rate function}. Let us now give the results for the rate function $\Phi({\sf e})$. We start by the typical value ${\sf e}_{\rm typ} = \Sigma'_{q=0} = f(\frac{\mu}{J}) - \log \mu $, from \eqref{etyp}, which gives more explicitly
\bea
&& {\sf e}_{\rm typ} = - \frac{1}{2} + \frac{\mu^2}{4 J^2} + \log \frac{J}{\mu} \quad , \quad \mu < 2 J. \\
&& {\sf e}_{\rm typ} = - \frac{1}{2} + \frac{\mu^2}{4 J^2} - \frac{\mu}{4 J} \sqrt{ \frac{\mu^2}{J^2} - 4} +
\log\frac{1 + \sqrt{1 - \frac{4 J^2}{\mu^2}}}{2}  \quad , \quad \mu > 2 J.
\eea

Let us consider first the case $\mu < 2J$.
For $q< 2 - \frac{\mu}{J}$ we can use the formula \eqref{complexSigmad0}
for $\Sigma_q$ in the complex phase.
We find that $q$ is determined by
\be
{\sf e} = \Sigma'_q = - \frac{1}{2} + \frac{\mu^2}{(2-q)^2 J^2} + \log \frac{J}{\mu}
\quad , \quad {\sf e} < {\sf e}_c = \frac{1}{2} + \log \frac{J}{\mu}, \label{ec}
\ee
where the last equation arises from the condition $q< 2 - \frac{\mu}{J}$ to
be able to use the complex phase formula \eqref{complexSigmad0}.
Let us define $\delta {\sf e} = {\sf e}- {\sf e}_{\rm typ}$. One finds
\be \label{above1}
\Phi({\sf e}) = \frac{\mu^2}{J^2} \phi( \frac{J^2}{\mu^2}  \delta {\sf e})\,,
\ee
where $\phi(x)$ is parametrically defined by eliminating $q$ in the system
\be  \label{param}
x = \frac{q(4-q)}{4 (2-q)^2} \quad , \quad \phi = \frac{q^2}{2(2-q)^2}\,,
\ee
which leads to
\be  \label{phi0simple}
\phi(x) = 1 + 2 x - \sqrt{1+ 4 x} = 2 x^2 - 4 x^3 + O(x^4)  \quad , \quad - \frac{1}{4} < x < x_c = \frac{J^2}{\mu^2} - \frac{1}{4}.
\ee
The limit $x \to - 1/4^+$ corresponds to $q \to - \infty$, but as we mentioned in the
introduction we restrict ourselves here to $q>-1$ (i.e. $x>-5/36$). The
region $x> x_c$ corresponds to ${\sf e}> {\sf e}_c$ and requires the use of the simple phase formula.
That formula is more cumbersome so we will not analyze it here. Similarly, the above
result \eqref{above1}, \eqref{phi0simple} remains correct in the region $\mu> 2 J$ for ${\sf e}> {\sf e}_c$, where ${\sf e}_c$
is still given by \eqref{ec} and is now smaller than ${\sf e}_{\rm typ}$, but it requires the use of the simple phase formula for ${\sf e}> {\sf e}_c$.

On the other hand we can also evaluate the rate function from the more general formula \eqref{resPhi}
\be
\Phi({\sf e}) =  \frac{1}{2} (\xi_q^*)^2 = \frac{1}{2} [ f^{-1} ( \delta {\sf e}  + f(\frac{\mu}{J}) ) - \frac{\mu}{J}  ]^2\,,
\ee
where we note that the function $f(z)$ can be shifted by any constant without changing the result.
This formula requires the evaluate the inverse function of $f(\xi)$, where $f(\xi)$ is given in \eqref{fzerod}.
We can see that this inversion is possible in closed form only for $|\xi|<2$.
In that case one can choose $f(z)=z^2/4$, which leads to the same result as \eqref{phi0simple}.
\\

\subsection{Explicit solution for $d=1$}.
\\

Here we will consider the continuum model for $d=1$. We set $t=1$ and $\Delta(k)=- k^2$. Note that for $d=1$ there is no need for an ultraviolet cutoff at large $k$.  This model is the one considered in \eqref{contmodel}
with $\tilde t=1$ and for convenience $\tilde J$ there is denoted here by $J$.

Recall that our notations in this section correspond to the study of the Hessian of elastic manifolds,
hence
we have set $E=0$ and used the parameter $\mu$. One may also be interested to compute
the generalised Lyapunov exponents defined in \eqref{Lambdaq}. They
can be obtained from $\Sigma_q$ computed below by the formula
\bea
&& \Lambda(q,E) = \left( \Sigma_q + \int \frac{dk}{2 \pi} (\log|k^2 + \mu| - \log(k^2)) \right)|_{\mu=-E}
 = \left( \Sigma_q + \sqrt{\mu} \theta(\mu) \right)|_{\mu=-E}
\eea
(where $\theta$ is the Heaviside step function)
since the subtraction for regularisation is slightly different in the two cases.
\\

{\it Larkin mass}.
We start with determining the Larkin mass $\mu_c$ in $d=1$. One has
\be \label{Larkin2}
1 = J^2  \int \frac{dk}{2 \pi}  \frac{1}{( \mu_c +  k^2 )^2}
= \frac{1}{4} J^2 \mu_c^{-3/2} \quad , \quad \mu_c= \left( \frac{J}{2} \right)^{4/3}.
\ee

{\it  Boundary for the continuous transition}. Let us determine $\mu_b$ given in \eqref{mubgen}.
One has
\be
\mu_b =  \mu_c - (q - 1) J^2 \int \frac{dk}{2 \pi}  \frac{1}{\mu_c + k^2} = \mu_c - (q-1) \frac{J^2}{2 \mu_c^{1/2}}
= (3- 2 q)  \left(\frac{J}{2}\right)^{4/3} = (3-2 q) \mu_c.
\ee
As expected, for $q=0$ the value of $\mu_b$ coincides (taking into account that $- \mu$ plays the role of the energy) with the edge of the density of states
$\mu_b(q \to 0)=- \alpha^*=3  (\frac{J}{2})^{4/3}$, see \ref{app:cubic}, which as discussed there and in
Section \ref{sub:connection} also coincides exactly with the transition point \eqref{Estar} of
the DBM in the cubic potential $\alpha^* = E^*= - 3  (\frac{J}{2})^{4/3}$. Note that
in the limit $q \to 0$ the saddle point is at $\xi_{q=0}^*=0$, hence the value of $c$ is immaterial
in that limit.
\\

{\it Simple phase.} The simple phase is $\mu> \mu_b=(3- 2 q)  (\frac{J}{2})^{4/3}$.
One has from the saddle point equation \eqref{solu2}, assuming $\mu_q>0$
\be
 \xi^*_q = q J \int \frac{d k}{2 \pi} \frac{1}{\mu_q  + k^2}
 = \frac{J q}{2 \mu_q^{1/2} }
 \quad , \quad \mu_q = \mu + (1- \frac{1}{q})J \xi^*_q.
\ee

This leads to a cubic equation for $\mu_q$
\be \label{eqmu}
\mu_q - \frac{J^2}{2}(q-1) \frac{1}{\mu_q^{1/2}} = \mu.
\ee
For $q=1$ one has $\mu_1=\mu$, For $q>1$ the l.h.s. is monotonically increasing
function of $\mu_q$ hence there is a unique root and $\mu_q$ is
an increasing function of $\mu$, with $\mu_q(\mu=0)=(2 (q-1))^{2/3} (J/2)^{4/3}$.
For $q<1$ the l.h.s of \eqref{eqmu} has a minimum at $\mu_q=\mu_q^* = (1-q)^{2/3} (J/2)^{4/3}$.
Hence for $\mu< 3 \mu_q^*$
there are no solutions, while for $\mu> 3 \mu_q^*$ there are
two roots $\mu_q^\pm$ with $\mu_q^- < \mu_q^* < \mu_q^+$. The correct root is the
largest one and, as expected, the branch $\mu_q^+$ corresponds
to the side $\mu>\mu_c$ for $q \to 0$ (since $\mu_0^*=\mu_c$). This result is
valid for $\mu> \mu_b = (3- 2 q)  (\frac{J}{2})^{4/3}$.
Note that for $\mu=\mu_b$ one has $\mu_q=\mu_c$.

To compute $\Sigma_q$ in the simple phase we use \eqref{dersigmu} which takes the form
\bea \label{dersigmu2}
&& \partial_\mu \Sigma_q =
\frac{1}{J} \xi^*_q -    \frac{q}{2}  \mu^{-1/2} = \frac{q}{2} ( \mu_q^{-1/2} - \mu^{-1/2}).
\eea
Changing integration variable from $\mu$ to
$\mu_q$ and using \eqref{eqmu}, gives
\bea \label{sigmu2}
&&  \Sigma_q = q \left(  \mu_q^{1/2} -  \mu^{1/2} - \frac{J^2 (q-1)}{8 \mu_q} \right),
\eea
where $\mu_q$ is the largest root of \eqref{eqmu}.
At the transition $\mu=\mu_b=(3-2 q) \mu_c$ one thus has, using that $\mu_q=\mu_c=(J/2)^{4/3}$,
for the rate of growth
\bea \label{sigmu3}
&&  \Sigma_q|_{\mu=\mu_b} = q (1- \sqrt{3-2 q} - \frac{q-1}{2} ) (J/2)^{2/3}.
\eea
One also
finds, by similar manipulations, the large deviation function
\be \label{2eqnew}
{\sf e} = \partial_q \Sigma_q = \mu_q^{1/2} -  \mu^{1/2} + \frac{J^2}{8 \mu_q}
\quad , \quad \Phi({\sf e}) = \frac{J^2 q^2}{8 \mu_q},
\ee
so that $\Phi({\sf e})$ can be obtained by eliminating
$q$ and $\mu_q$ in the three equations \eqref{2eqnew}
and \eqref{eqmu}. This is valid for ${\sf e}< {\sf e}_c$ where
\be
{\sf e}_c = \mu_c^{1/2} - \mu^{1/2} + \frac{J^2}{8 \mu_c} = \frac{3}{2} (J/2)^{2/3}  - \mu^{1/2}.
\ee
For ${\sf e}>{\sf e}_c$ one needs the formula from the complex phase, addressed below.

Let us give the value of ${\sf e}_{\rm typ}$ in more details.
From the above one has, in the simple phase $\mu > \mu_b(q=0)=3 (J/2)^{4/3}$
\be
{\sf e}_{\rm typ} =  (J/2)^{2/3} (\tilde \mu_0^{1/2} -  \tilde \mu^{1/2} + \frac{1}{2 \tilde \mu_0} )  \quad , \quad
\tilde \mu_0 +  \frac{2}{\tilde \mu_0^{1/2}} = \tilde  \mu\,,
\ee
where $\mu = \tilde \mu (J/2)^{4/3}$ and
$\tilde \mu_0$ being the largest root of the second equation.
On the other hand from \eqref{etyp} one also has
\be
{\sf e}_{\rm typ} = f(\mu/J) - \int \frac{dk}{2 \pi} \log(k^2 + \mu)
= \int d\alpha (\rho_K(\alpha) - \rho_0(\alpha) ) \log| \alpha + \mu|\,,
\ee
where $\rho_0(\alpha)= \int \frac{dk}{2 \pi}  \delta(\alpha-k^2)= 1/(2 \pi \sqrt{\alpha}) \theta(\alpha)$ denotes the free DOS and the substraction ensures the convergence at large $\alpha$.
Let us recall the result of Ref \cite{UsHessianManifold} for $\rho_K(\alpha)$ in the more convenient form
derived in present paper, see \eqref{newDOS}
\be
\rho_K(\alpha)=
\frac{\sqrt{3}}{4 \pi (2 J^2)^{1/3}} g(\Lambda=\frac{\alpha}{3 (J/2)^{4/3}}) \quad , \quad g(\Lambda)=
 (1 + \sqrt{1 + \Lambda^3})^{2/3}  - (1 - \sqrt{1 + \Lambda^3})^{2/3}.
\ee
This gives
\be
{\sf e}_{\rm typ} = \frac{\sqrt{3}}{2 \pi} (\frac{J}{2})^{2/3}
\int_{-1}^{+\infty} d\Lambda ( \frac{3}{4} g(\Lambda)- \frac{1}{\sqrt{\Lambda}} \theta(\Lambda) )
\log(\tilde  \mu + 3  \Lambda)   \quad , \quad \tilde \mu = \mu/(J/2)^{4/3}
\ee
where we used that $\int_{-1}^{+\infty} d\Lambda ( \frac{3}{4} g(\Lambda)- \frac{1}{\sqrt{\Lambda}} \theta(\Lambda) ) =0$. Since it is not at all obvious, we have checked numerically
that this gives the same value for ${\sf e}_{\rm typ}$, showing consistency of
our approach. In the simple
phase one has ${\sf e}_{\rm typ} < {\sf e}_c$ and
one finds that for $\mu \to \mu_b(q=0)=3  (J/2)^{4/3}$ one has
${\sf e}_{\rm typ}  \to {\sf e}_c= (\frac{3}{2} - \sqrt{3}) (J/2)^{4/3}$.
\\

{\it Complex phase}. In that phase the equations are more
complicated so we only sketch how one can compute $\Sigma_q$.
The saddle point equations \eqref{solu3}
and \eqref{solu4} for the complex phase become in $d=1$
\be \label{spc1}
 1= J^2 \int \frac{d k}{2 \pi} \frac{1}{( \mu_q + k^2)^2 + J^2 y^2}   \quad , \quad J \tilde \xi^*_q = J^2 q
\int \frac{d k}{2 \pi} \frac{\mu_q + k^2}{( \mu_q + k^2)^2 + J^2 y^2}  \quad , \quad \mu_q = \mu + (1- \frac{1}{q})J \xi^*_q.
\ee
Let us define the two integrals
\be
I_1(\hat y) = \int_0^{+\infty} \frac{dz}{2 \pi\sqrt{z}} \frac{1}{(1+ z)^2 + \hat y^2} \quad , \quad
I_2(\hat y) = \int_0^{+\infty} \frac{dz}{2 \pi\sqrt{z}} \frac{1+z}{(1+ z)^2 + \hat y^2}\,,
\ee
which are easily computed using the identities
\be
I_2(\hat y) \mp  i  \hat y I_1(\hat y) = \int_0^{+\infty} \frac{dz}{2 \pi\sqrt{z}} \frac{1}{1+ z \pm i \hat y} = \frac{1}{2 \sqrt{1 \pm  i \hat y} }.
\ee
Assuming $\mu_q>0$ the saddle point equations \eqref{spc1} can be rewritten as
\be
1 = J^2 \mu_q^{-3/2} I_1(\hat y = \frac{J y}{\mu_q}) \quad , \quad
J \xi = J^2 q \mu_q^{-1/2} I_2(\hat y = \frac{J y}{\mu_q}).
\ee
The saddle point equations can be rewritten as
\be
\xi_q^* = q \frac{\mu_q-\mu}{q-1} \quad , \quad
\frac{1}{J^2} ( \frac{\mu_q-\mu}{q-1} \mu_q^{1/2} - i \hat y \mu_q^{3/2} ) = \frac{1}{2 \sqrt{1+ i \hat y} } \quad , \quad
\hat y =  \frac{J y}{\mu_q}.
\ee
Eliminating $\hat y$ it determines $\mu_q$, hence $\xi_q^*$
as a function of $\mu$.
From there one can access $\Sigma_q$ using the relation
$\partial_\mu \Sigma_q =
\frac{1}{J} \xi^*_q -    \frac{q}{2}  \mu^{-1/2}$
and its known value \eqref{sigmu3} at $\mu=\mu_b=(3-2 q) (J/2)^{4/3}$.

\bigskip

{\bf Acknowledgements} We thank B. Rider for pointing Ref. \cite{BaurKratz} to us.
YVF and PLD gratefully acknowledge the hospitality of the programme "Lush World of Random Matrices" at ZiF, University of Bielefeld,
during the final stages of this project. AO acknowledges support by  EPSRC grant  EP/V002473/1 "Random Hessians and Jacobians: theory and applications"
and YFV by  UKRI1015 "Non-Hermitian random matrices: theory and applications".
PLD acknowledges support from ANR grant ANR-17-CE30-0027-01 RaMaTraF.

\appendix

\renewcommand\thesection{Appendix \Alph{section}.}
\renewcommand\thesubsection{\Alph{section}.\arabic{subsection}}
\renewcommand\theremark{\Alph{section}.\arabic{remark}}

\section{Reminder on the matrix Ricatti equation and the Lyapunov exponents}
\label{app:ricatti}

Let us consider the Schr\"odinger equation for a
vector $\psi(\tau)=\{ \psi_j(\tau) \}_{j=1,\dots,N}$ (we set $t=1$)
\bea \label{model2}
\psi''(\tau) = (W(\tau) - E \mathbf{1}) \cdot \psi(\tau)\,,
\eea
which we rewrite as
\be
\psi'_i(\tau) = \phi_i(\tau) \quad , \quad \phi'_i(\tau) = (W(\tau) - E \mathbf{1})_{ij}  \psi_j(\tau).
\ee
Following Refs.
\cite{GrabschThesis,GT16}
we define $\Psi(\tau),\Phi(\tau)$ two matrices made of $N$ independent solutions
$k=1,\dots,N$ of
this equation hence we have
\be
\Psi'_{ik}(\tau) = \Phi_{ik}(\tau) \quad , \quad \Phi'_{ik}(\tau) = (W(\tau) - E \mathbf{1})_{ij} \Psi_{jk}(\tau).
\ee
Let us take for definiteness a set of $N$ initial conditions, $\Psi(0)=0$ and $\Phi(0)=\Psi'(0)=\mathbf{1}$,
that is the $k$-th solution has $\psi_i(0)=0$ and $\psi_i'(0)=\delta_{ik}$.

We now define the matrix
\be
{\cal Z}(\tau) = \Phi(\tau) \Psi(\tau)^{-1} = \Psi'(\tau) \Psi(\tau)^{-1}.
\ee
Then ${\cal Z}(\tau)$ satisfies the matrix Ricatti equation
\bea
&& d {\cal Z}(\tau)  = d \Phi(\tau) \Psi(\tau)^{-1}  - \Phi(\tau) \Psi(\tau)^{-1} d\Psi(\tau) \Psi(\tau)^{-1} \\
&& = (W(\tau) - E \mathbf{1}) \Psi(\tau) \Psi(\tau)^{-1} d\tau - \Phi(\tau) \Psi(\tau)^{-1} \Phi(\tau) \Psi(\tau)^{-1} d\tau \\
&& = [W(\tau) - E \mathbf{1} - {\cal Z}^2] d\tau.
\eea

Note that we have
\be
{\rm Tr} {\cal Z}(\tau) = {\rm Tr} \Psi'(\tau) \Psi(\tau)^{-1} = \frac{d}{d\tau} {\rm Tr} \log \Psi(\tau).
\ee
Hence
\be
\int_0^L d\tau {\rm Tr} {\cal Z}(\tau) = \log \frac{\det \Psi(\tau)}{\det \Psi(0)}\,,
\ee
which is equal to the regularized $\log |\det({\cal H} - E)|$ by generalisation of
Gelfand-Yaglom relation derived in \cite{OssipovYaglom}.

To relate to Lyapunov exponents let us recall that, see e.g. \cite{GrabschThesis},
the sum of the $1 \leq n \leq N$ largest Lyapunov exponents is defined as
\be
\lim_{\tau \to +\infty} \frac{1}{2\tau} \log |\Psi^+ \Psi|_n = \sum_{i=1}^n \gamma_i\,,
\ee
where $|M|_n$ is the determinant of the $n \times n$ minor (top left subblock) of $M$.
Hence if one considers $n=N$ one obtains
\be
{\rm Tr} {\cal Z}(\tau) = \sum_{n=1}^N \hat \gamma_n.
\ee

\section{Evolution of the resolvent}  \label{app:resolvant}

Having defined $G(z,\tau) = \frac{1}{N} \sum_i \frac{1}{\lambda_i(\tau)-z}$
and using Ito's rule, the stochastic equation \eqref{Langevin1} leads to
\be
dG(z,\tau) =  \frac{1}{N} \sum_i d\lambda_i \, \partial_{\lambda_i} \frac{1}{\lambda_i-z} +
\frac{\tilde J^2}{N} d\tau  \sum_i \partial_{\lambda_i}^2 \frac{1}{\lambda_i-z}.
\ee
After standard manipulations, i.e. $\partial_{\lambda_i} \equiv - \partial_z$,
$\frac{1}{N} \sum_{i \neq j} \frac{1}{\lambda_i-\lambda_j} \frac{1}{z-\lambda_i}
= - \frac{1}{2} ( N G(z,\tau)^2 + \partial_z G(z,\tau))$,
as well as $\frac{1}{N} \sum_i \lambda_i^2 \partial_{\lambda_i} \frac{1}{\lambda_i-z}
= \partial_z( z + z^2 G(z,\tau))$,
one obtains \eqref{eq5} in the text,
where $\hat \eta(z,\tau)= - \sqrt{\frac{2}{N}} \sum_i \frac{dB_i}{d\tau (\lambda_i-z)}$
is a Gaussian noise with correlator given by \eqref{eq6} in the text.

\section{Relation between the Dyson Brownian motion in a cubic potential
and the DOS of a matrix valued random Schr\"odinger operator} \label{app:cubic}

Below we sketch some details of the Allez and Dumaz calculation \cite{AD15} and
provide the explicit comparison with our result for the density of state $\rho_K(\alpha)$
of the matrix Schr\"odinger operator $K_{ij}(\tau) = - \tilde t \partial_\tau^2 \delta_{ij} + \tilde J \tilde H_{ij}(\tau)$.

\subsection{Density of states from the saddle point method}

Let us first recall the explicit formula obtained by two of us in \cite{UsHessianManifold}
for the density of states of the matrix
${\cal H}$ in \eqref{discretemodel}, with the parameter $J^2 = 4 B''(0)$ and $c=0$
(a more general problem was solved there, but we restrict to that
situation which amounts to set $\mu_{\rm eff}=0$ there). On page 196,
example 3, we considered the model in $d=1$
in the continuum limit with $- t \Delta(k) = t k^2$. Setting $t=1$
it is thus the same as the matrix Schr\"odinger operator $K$.
We found that
$\rho_K(\alpha)= \frac{1}{\pi} {\rm Im} (i p)$ where $p$ is a complex
parameter satisfying the self-consistent equation (Eq. (78) there)
\bea
i p = \int \frac{dk}{2 \pi} \frac{1}{\alpha - t k^2 - 4 i p B''(0)}.
\eea
Note that $p$ should be identified with $i r_\lambda$ in Section \eqref{sec:part2}.
Performing the integral one gets a cubic equation for $p$. Solving it
leads to (recalling that for the continuum model $B''(0)=\tilde J^2/4$)
\bea \label{resN}
&& \rho_K(\alpha) = \frac{1}{2 \pi ({\tilde J}/2)^{2/3}} r_c(\frac{\alpha}{3 ({\tilde J}/2)^{4/3}}) \quad , \quad r_c(\Lambda)= \frac{w_r^2}{4}\,,
\sqrt{ (\frac{2}{w_r})^3 -1 } \\
&&  w_r = (1 + \sqrt{1+ \Lambda^3})^{1/3} +
 (1 - \sqrt{1+ \Lambda^3})^{1/3} \quad , \quad \Lambda > -1. \\
 &&  \rho_K(\alpha) =  0 \quad , \quad \Lambda < -1.
\eea
where $(1 - \sqrt{1+ \Lambda^3})^{1/3}= {\rm sgn}(1 - \sqrt{1+ \Lambda^3})|1 - \sqrt{1+ \Lambda^3}|^{1/3}$.
Hence the left spectral edge is at $\alpha=\alpha^* = 3 ({\tilde J}/2)^{4/3}$, where
the density  $\rho_K(\alpha)$ vanishes as a square root.
For large $\alpha$ it vanishes as $\rho_K(\alpha) \simeq 1/(2 \pi \sqrt{\alpha})$.

\subsection{Comparison with the Allez-Dumaz calculation}

It is useful to recall the correspondence of our parameters with \cite{AD15}:
\be \label{correspondence}
a =-E \quad , \quad \beta = 2  {\tilde J}^2.
\ee
We will use both notations below. To study the solution \eqref{Gpm} for the stationary resolvent
and to find the value of the integration constant ${\cal J}$ as well as the branch which
leads to the proper solution (corresponding to a non-negative eigenvalue density $\rho(\lambda)$ for the
Ricatti matrix $Z$)
Allez and Dumaz introduce the polynomial
\be
P(z) = (z^2 + E)^2 - 2 {\tilde J}^2 (z - {\cal J})
\ee
and show that it must not have any root with odd multiplicity in $U$  (the strict upper complex plane)
since we want $G(z)$ to be analytic in $U$. Hence $P(z)$ and $P'(z)$ must vanish for the same $z$. On the other hand from Cardano's formula $P'(z) = 4 z^3 + 4 E z - 2 {\tilde J}^2$ has 3 real roots if and only if
\bea
- E > \frac{3}{4} (2 {\tilde J}^2)^{2/3} = - E^* = 3 ({\tilde J}/2)^{4/3}.
\eea
There is a phase transition at this value of $E$ and we see that it corresponds exactly to the edge of the spectrum
of $\rho_K(\alpha)$ in \eqref{resN}, i.e. $E^*=\alpha^*$.

To satisfy the constraint mentioned above one must have
\be
P(z) = (z-z_a)^2 (z-\gamma_-)(z-\gamma_+)\,,
\ee
where the double root $z_a$ satisfies $P'(z_a)=0$,
and the two other roots are given by
\be
\gamma_\pm = - z_a \pm \sqrt{2(a-z_a^2)} \,,
\ee
which is valid for any $a$.
The double root $z_a$ is

(i) real for $a>a^*$, where $a^*= \frac{3}{4} \beta^{2/3}$,
i.e. $E < E^*$, that is outside the spectrum of $K$ and

(ii) has ${\rm Im} z_a>0$ for $a<a^*$, i.e. $E > E^*$, that is inside the spectrum of $K$.
In that case it reads (correcting a misprint in \cite{AD15})
\be
z_a = \frac{\beta^{1/3}}{2} (( - \frac{1}{2} + \frac{\sqrt{3}}{2} i) (1 + \sqrt{1 - (a/a^*)^3})^{1/3}
 - (\frac{1}{2} + \frac{\sqrt{3}}{2} i) {\rm sgn}(1 - \sqrt{1 - (a/a^*)^3}) |1 - \sqrt{1 - (a/a^*)^3}|^{1/3} ).
\ee

The condition that the root $z_a$ is double determines the integration constant ${\cal J}$ since
one must have
\be \label{2eq0}
P(z_a)=(z_a^2 -a)^2 - \beta (z_a - {\cal J}) = 0 \quad , \quad P'(z_a)= 4 z_a (z_a^2 -a)  - \beta = 0.
\ee
Hence the integration constant is given by
\be
{\cal J} = z_a - \frac{1}{\beta}  (z_a^2 -a)^2 = z_a - \frac{\beta}{(4 z_a)^2}
\ee
which is

(i) real for $a>a^*$, i.e. $E < E^*$, that is outside the spectrum of $K$ and

(ii) has ${\rm Im} {\cal J} >0$ for $a<a^*$, i.e. $E > E^*$, that is inside the spectrum of $K$.
\\

We can now test the prediction in \eqref{conjecture} which we rewrite here
\be \label{conjecture2}
\rho_K(\alpha) = \frac{1}{\pi}  \frac{d}{dE} {\rm Im} {\cal J}(E)|_{E=\alpha}.
\ee
One has
\bea
\frac{1}{\pi}  \frac{d}{dE} {\rm Im} {\cal J}(E) =
- \frac{1}{\pi}  \frac{d}{da} {\rm Im} {\cal J} =  - \frac{2}{\pi \beta} {\rm Im} z_a^2\,,
\eea
where the last equality is obtained from taking a derivative w.r.t. $a$ of
the equation $P(z_a)=0$, see \eqref{2eq0}, and using that $P'(z_a)=0$
which gives
\be
\beta \frac{d {\cal J}}{da}
= 2 z_a^2 - 2 a,
\ee
recalling that $a=-E$ is real.
Now, denoting $a/a^* = - \Lambda$ we find that
\be
z_a = \frac{\beta^{1/3}}{2} (( - \frac{1}{2} + \frac{\sqrt{3}}{2} i) (1 + \sqrt{1 + \Lambda^3})^{1/3}
 - (\frac{1}{2} + \frac{\sqrt{3}}{2} i) {\rm sgn}(1 - \sqrt{1 + \Lambda^3}) |1 - \sqrt{1 + \Lambda^3 }|^{1/3} )
\ee
which leads to
\be \label{newDOS}
\frac{1}{\pi}  \frac{d}{dE} {\rm Im} {\cal J}(E) = \frac{\sqrt{3}}{4 \pi \beta^{1/3}}
( (1 + \sqrt{1 + \Lambda^3})^{2/3}  - (1 - \sqrt{1 + \Lambda^3})^{2/3} ).
\ee
We find that this expression is equivalent for $\Lambda > -1$ to
the expression given in \eqref{resN} for $\rho_K(\alpha)$.

\subsection{Barrier crossing probability}

Let us first recall the main results of Ref. \cite{AD15}. There are two phases.
The density in each phase is as follows:\\

{\bf Phase with current: $a \leq a^*$}.
If $a \leq a^*$ then ${\rm Im} z_a >0$ and ${\rm Im} \gamma_\pm <0$. The support of the
density of the $\lambda_i$'s
extends to the full real axis, and the density and its tail are given by
\be
\rho(\lambda) = \frac{2}{\pi} {\rm Im} \sqrt{P(\lambda)} \quad , \quad
\rho(\lambda) \simeq_{\lambda \to \pm \infty} \frac{1}{\pi} \frac{{\rm Im} {\cal J} }{\lambda^2}.
\ee
For the Schr\"odinger operator it corresponds to being inside the spectrum with $\rho_K(\alpha)>0$.
\\

{\bf Confined zero-current phase: $a \geq a^*$}.
If $a > a^*$, one finds that the roots of $P$ and $P'$ are real with $z_a < \gamma_- < \gamma_+$.
The support of the density is bounded and equal to the interval $[\gamma_-,\gamma_+]$.
Inside its support the density reads
\bea \label{rhoconfined}
\rho(\lambda) = \frac{2}{\beta \pi}  \sqrt{- P(\lambda)} = \frac{2}{\beta \pi}
\sqrt{\beta(\lambda - {\cal J} ) - (\lambda^2 - a)^2} = \frac{2}{\beta \pi} (\lambda-z_a) \sqrt{(\lambda-\gamma_-)(\gamma_+-\lambda)}.
\eea
In this phase ${\cal J}$ is real so there is no current of particles.
For the Schr\"odinger operator it corresponds to being outside the spectrum with $\rho_K(\alpha)>0$.
\\

Finally, at the transition $a=a^*$ the support of the density is the interval
$[\gamma_-,\gamma_+]$ with $\gamma_-= z_a=
-\frac{1}{2} \beta^{1/3}$ and $\gamma_+=\frac{3}{2} \beta^{1/3}$, and inside its support it reads
\bea
\rho(\lambda) = \frac{2}{\beta \pi} (\lambda - \gamma_- )^{3/2} \sqrt{\gamma_+ - \lambda}.
\eea
 The exponent $3/2$ obtained in \cite{AD15} is a distinct universality class from the usual semi-circle edge behavior.
\\

{\bf Barrier crossing}.
We now study the probability of barrier crossing in the confined phase.
For this we consider the left-most eigenvalue, call it $\lambda_1$, and allow
its position to vary to the left of the support so it can eventually escape to $-\infty$.
From the equation of motion \eqref{Langevin1}
we have the Langevin equation (this is the
model with $c=0$)
\be
\frac{d \lambda_1}{d\tau} = (a- \lambda_1^2) - \frac{\beta}{2} G_N(\lambda_1)
+ \frac{\sqrt{2 \tilde J^2}}{\sqrt{N}} \eta_i(\tau)\,,
\ee
where we have used the definition of the resolvent $G_N(z)$ at finite $N$.
Until now this is exact.

We now study the large $N$ limit. We will keep the Brownian noise (since
it allows for barrier crossing) but approximate $G_N(z)$
by its infinite $N$ limit, $G(z)$, i.e. we assume that the density
of the other eigenvalues, $\rho(\lambda)$, is
still given by \eqref{rhoconfined} (for similar type of large deviation arguments
in the context of the standard DBM see e.g. \cite{MaidaLargeDev}).
The deterministic "force" $f(\lambda_1)$ which acts on the
particle $\lambda_1$ is thus, using the result of \cite{AD15} for $G(z)$ (see \eqref{Gpm} with
the correct branch) after recalling the correspondence \eqref{correspondence}
\bea
f(\lambda_1) = a- \lambda_1^2 - \frac{\beta}{2} G(\lambda_1)
= \sqrt{ P(\lambda_1)} = (\lambda_1-z_a) \sqrt{(\gamma_- - \lambda_1) (\gamma_+ - \lambda_1)}
\eea
for $\lambda_1> \gamma_-$.
We note that this force is the sum of the force from the cubic potential
and the repulsive force from the other eigenvalues. We see that
it precisely vanishes at $\lambda_1 \to \gamma_-$ , i.e. when
$\lambda_1$ reaches the left edge of the spectrum, as it should
since this is a global equilibrium. However we see that it
also vanishes at the point $\lambda_1=z_a < \gamma_-$, which is the top of the barrier,
and
that in the interval $[z_a , \gamma_-]$ it is strictly positive, i.e.
the particle is pushed to the right. The barrier
crossing potential energy is thus
\be
U = \int_{z_a}^{\gamma_-}  d\lambda (\lambda - z_a) \sqrt{( \gamma_--\lambda ) ( \gamma_+- \lambda) }.
\ee
Computing the integral and using $\gamma_\pm= -z_a \pm \sqrt{2 (a-z_a^2)}$ as
well as $P'(z_a)=0$ we obtain
\bea \label{barrierexact}
U = \frac{2}{3} a \sqrt{6 z_a^2-2 a}
-\beta  \sinh^{-1}\left(\frac{\sqrt{\frac{4 \left(-z_a\right){}^{3/2}}{\sqrt{\beta}}-\sqrt{2}}}{2^{3/4}}\right).
\eea
Near the transition we find
\be
U \simeq \frac{4}{5} \sqrt{2} \times 3^{1/4} \beta^{1/6}  \,  (a-a^*)^{5/4}.
\ee
We have used  $z_a \simeq - \beta^{1/3} ( \frac{1}{2} + \frac{\sqrt{a-a^*}}{\sqrt{3}} - \frac{2}{3} (a-a^*))$,
$\gamma_- - z_a \simeq  \sqrt{3} \sqrt{a-a^*}$, and that in this region
one can approximate
\be
U \simeq (\gamma_+-\gamma_-)^{1/2} \int_{z_a}^{\gamma_-} d\lambda (\lambda - z_a) \sqrt{\gamma_--\lambda}
= \frac{4}{15} (\gamma_+-\gamma_-)^{1/2} (\gamma_--z_a)^{5/2}.
\ee
In the limit $a = - E \to + \infty$ we obtain $z_a \simeq - \sqrt{a}$
and $\gamma_\pm \simeq \sqrt{a}$,
which leads to
\be
U \simeq \frac{2}{3} a \sqrt{4 a} = \frac{4}{3} |E|^{3/2}.
\ee

Since the temperature here is $T={\tilde J}^2/N$ we see that the Kramers crossing time is
\bea
\tau_{\rm crossing} \sim \exp(  \frac{N U}{{\tilde J}^2} ).
\eea
Hence there will be typically at least one DBM particle going over
the barrier in that time scale. The counting theorems then imply
that the distance between "nodes" is of order $\tau_{\rm crossing}$
hence that the DOS of the Schrodinger operator (equivalently
the DBM particle current) is of order
\bea
\rho_K(\alpha) \sim \tau_{\rm crossing}^{-1} \sim \exp( -  \frac{N U}{{\tilde J}^2} ) \quad , \quad U=U|_{a=-\alpha}.
\eea

\end{document}